# Pulse compression by parametric beating with a prepared Raman coherence

Fam Le Kien[*] and K. Hakuta
*Department of Applied Physics and Chemistry, University of Electro-Communications, Chofu, Tokyo 182-8585, Japan*
*CREST, Japan Science and Technology Corporation (JST), Chofu, Tokyo 182-8585, Japan*

A. V. Sokolov
*Department of Physics and Institute for Quantum Studies,*
*Texas A&M University, College Station, TX 77843-4242*

(Dated: July 17, 2002)

We present a general analysis for the interaction of a probe laser radiation with a coherently prepared molecular Raman medium. We describe a general formalism which includes dispersive effects, such as group velocity and group velocity dispersion (GVD). When dispersion is negligible, the analysis is especially simple and insightful. We show that molecular oscillations result in a modulated instantaneous susceptibility of the medium. The effect of the time-varying susceptibility on a probe laser pulse is two-fold: the output frequency becomes modulated because of the time-varying phase velocity, and the pulse shape becomes deformed because of the time-varying group velocity. We identify two mechanisms for pulse compression: (1) Frequency chirping with subsequent pulse compression by normal linear GVD (possibly in the same medium) and (2) Compression due to the time-varying group velocity. We analyze various aspects of pulse compression in the coherent Raman medium and derive conservation relations for this process. When we consider a probe laser pulse which is much shorter than the molecular oscillation period, we observe frequency chirping, compression, or stretching of this pulse, depending on its relative timing with respect to the molecular oscillations. Based on our analysis, we propose a method for selective compression or frequency conversion of single ultrashort pulses.

## I. INTRODUCTION

Generation of ultrashort pulses is a rapidly developing and highly motivated field of optics [1]. Until recently the shortest optical pulses were produced by solid-state Ti:Sapphire laser systems. Pulses as short as 4.5 fs (just under two optical cycles) were obtained by expanding the spectrum of a mode-locked laser by self-phase modulation in an optical waveguide, and then compensating for group velocity dispersion by diffraction grating and prism pairs [2]. The development of solid-state femtosecond laser technology allowed time-resolved studies of chemical reactions and molecular dynamics [3]. Generation of even shorter pulses would extend the horizon of ultrafast measurements to the time scale of electronic motion [4], but the progress seemed to be stuck at the few-femtosecond barrier for many years.

The year 2001 saw a breakthrough in ultrashort pulse generation in several directions simultaneously. Workers in the field of high-order harmonic generation have measured subfemtosecond pulses in the X-ray spectral region [5]. Impulsive Raman scattering has produced pulses as short as 3.8 fs in the near ultraviolet [6]. The adiabatic Raman technique has been used to demonstrate collinear generation of a wide spectrum of mutually coherent sidebands (spanning infrared, visible, and UV regions) [7, 8] that can synthesize pulses as short as one optical oscillation ($\sim 2$ fs) [9], and even sub-cycle pulses. These different techniques can be considered as complimentary to each other, as they provide ultrashort pulses with very different characteristics. High-order harmonic generation is a unique source of X-ray pulses, but by their very nature these pulses are difficult to control because of intrinsic problems of X-ray optics; besides the conversion efficiency into these pulses is very small (typically $10^{-5}$) [10]. On the other hand, the adiabatic Raman technique allows 100% conversion [11] and produces a well-controlled spectrum centered in the visible region. A disadvantage of the adiabatic technique is that it leads to trains of many pulses. In the original configuration [12, 13] this technique produces trains of pulses spaced by the molecular oscillation period, which is as short as 11 fs for molecular vibrations in deuterium.

Several ideas have been put forward for generation of single subfemtosecond pulses by the Raman techniques. Early work on the dynamics of a single intense laser pulse in a Raman-active medium included predictions of $2\pi$ soliton formation and subfemtosecond pulse compression [14]. Recent experiments have shown that a weak probe pulse can be compressed by molecular oscillations which are excited impulsively by a strong pump pulse [6, 15, 16]. Kalosha *et al.* have suggested that molecular wavepacket revivals can produce frequency chirp, which in turn would allow femtosecond pulse compression by normal group velocity dispersion (GVD) in a thin output window [17]. Bartels *et al.* have demonstrated this possibility exper-imentally [18]. A disadvantage of the impulsive excitation method compared to the adiabatic technique is that the excitation level and the generated Raman coherence are several orders of magnitude smaller. Therefore, the impulsive excitation method requires strong pump intensity, for which nonlinear response of the medium becomes substantial and ultimately limits the level of excitation. There has been a proposal [19] to combine the adiabatic technique with the impulsive technique. According to this method, a short probe pulse can be compressed into a singlet, doublet or triplet of subfemtosecond pulses by beating with an adiabatically prepared Raman coherence. This method has the advantage of the adiabatic pumping in producing a large Raman coherence and the advantage of the impulsive excitation in reducing the number of pulses per train.

---

[*]On leave from Department of Physics, University of Hanoi, Hanoi, Vietnam; also at Institute of Physics, National Center for Natural Sciences and Technology, Hanoi, Vietnam

In this paper we again consider the combination of the adiabatic and impulsive techniques. We perform an adiabatic Raman excitation by two narrow-linewidth laser fields, slightly detuned from the Raman resonance. We then apply a single ultrashort pulse. We show that, when the pulse is short compared to the molecular half-period, it can be stretched, compressed, or frequency converted. In essence, coherent molecular motion results in a time-varying instantaneous susceptibility, which in turn causes variations in phase and group velocities and produces a time-varying gain coefficient. The use of the terminology such as the time-varying susceptibility can be justified when dispersion is small and the electronic response can be treated as instantaneous [20]. Time-varying phase velocity leads to frequency modulation and results in an up- or down-conversion of the probe laser pulse, or in a pulse chirp, depending on the pulse timing with respect to the molecular motion. Chirped pulses are then either stretched or compressed, depending on the sign of the chirp, by normal GVD, possibly in the same medium [12]. The second mechanism for pulse compression or stretching is the direct action of the time-varying group velocity [16, 19]. These mechanisms for frequency modulation and pulse compression are reminiscent of self-phase modulation [21], self-modulation of plasma wake fields [22], and self-steepening [23], which are produced by self-induced nonlinear susceptibility. Similar to self-steepening [23], our second mechanism comes into play when the time scale for significant variations of susceptibility becomes comparable to the duration of an optical cycle.

It has been shown earlier [12, 13] that, in a general case, with all dispersive effects included, the number of photons in the laser field interacting with a Raman system, is conserved. In this paper we obtain additional conservation relations for the limit of negligible dispersion. We derive the conservation of the area of a laser pulse interacting with coherently prepared molecular oscillations [24]. This leads to a possibility of increased energy accompanying frequency up-conversion. In this situation, the energy added to the field is extracted from the internal energy of the medium. We also find that, as the pulse length, frequency, amplitude, and energy change during the propagation in the medium, in addition to the photon number and pulse area conservations, the number of optical oscillations as well as the product of the pulse length and the mean frequency are conserved. Even though these conservation relations are derived in the idealized case, they are useful for the general analysis of the problem. Based on this analysis, we propose a method for generation of single subfemtosecond pulses. We suggest applying a pulse short compared to the molecular half-period. Then the output will depend on the timing between the pulse and the independently prepared molecular motion. In principle, this timing can be controlled by phase-locking among lasers used to prepare and to probe the molecular coherence. In practice, this would be a very challenging experimental task. Alternatively one can assume random shot-to-shot timing, and use a sorting algorithm (based on spectral analysis) for the output pulse selection.

Before we proceed, we note that there is a large history of ideas for short pulse manipulation by the Raman techniques. Ruhman *et al.* have used impulsive stimulated Raman scattering to observe coherent molecular vibrations in the time domain, and discussed mechanisms for frequency conversion [25]. Imasaka and colleagues have observed generation of a broad rotational Raman spectrum in molecular hydrogen, and discussed possibilities for phase-locking this spectrum [26]. Kaplan and Shkolnikov have predicted the existence of $2\pi$ Raman solitons with a phase-locked spectrum that Fourier-transforms into a train of subfemtosecond pulses [27]. In other related work Kocharovskaya *et al.* have suggested using a Raman medium inside a laser cavity so as to provide phase modulation to cause mode locking [28].

The paper is organized as follows. In Sec. II we describe the model and present the basic equations. In Sec. III we analyze the compression of a probe field due to the beating with a prepared Raman coherence. Finally, Sec. IV contains conclusions.

## II. MODEL AND BASIC EQUATIONS

We consider the interaction of the electromagnetic field with a molecular medium whose levels and transitions are shown schematically in Fig. 1. Levels $j$ with energies $\omega_j$ are coupled to level $a$ with energy $\omega_a$ and level $b$ with energy $\omega_b$ by electric dipole transitions. The transition between levels $a$ and $b$ is electric dipole forbidden. We allow an arbitrary number of levels $j$ and analyze the system by including all possible components of the Raman spectrum. We assume that the transitions $j \leftrightarrow a$ and $j \leftrightarrow b$ are far off one-photon resonance, the transition $a \leftrightarrow b$ may be off two-photon resonance by a detuning $\delta$, the field is linearly polarized, and the propagation is one-dimensional, along the $z$ direction.

In the dipole approximation, the Hamiltonian of the

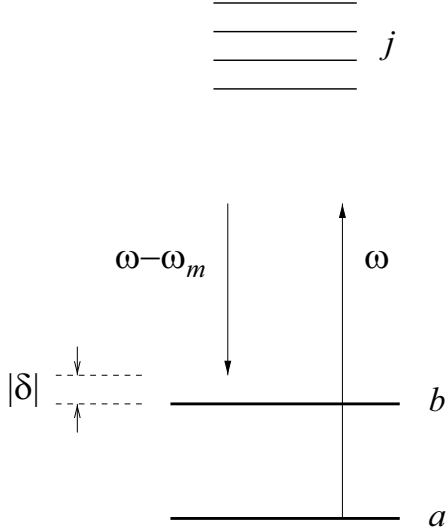

FIG. 1: Diagram of energy levels and transitions for the analysis. Levels $a$ and $b$ are coupled with levels $j$ by the spectral components $\omega$ and $\omega - \omega_m$ of the fields. The transition between levels $a$ and $b$ is electric dipole forbidden. The transitions $j \leftrightarrow a$ and $j \leftrightarrow b$ are far off one-photon resonance, and the transition $a \leftrightarrow b$ may be off two-photon resonance by a detuning $\delta$.

interaction of the field with a molecule is given by

$$H = H_0 + H_{\text{int}}, \tag{1}$$

where

$$H_0 = \hbar\omega_a \sigma_{aa} + \hbar\omega_b \sigma_{bb} + \sum_j \hbar\omega_j \sigma_{jj} \tag{2}$$

and

$$H_{\text{int}} = -\sum_j E(\mu_{ja}\sigma_{ja} + \mu_{aj}\sigma_{aj} + \mu_{jb}\sigma_{jb} + \mu_{bj}\sigma_{bj}). \tag{3}$$

Here $\sigma_{ik} = |i\rangle\langle k|$ are the operators for energy level populations ($i = k$) and transition amplitudes ($i \neq k$), $E$ is the electric field, and $\mu_{ja}$ and $\mu_{jb}$ are the dipole moments of the transitions $j \leftrightarrow a$ and $j \leftrightarrow b$, respectively. For simplicity, we assume that $\mu_{aj}$ and $\mu_{bj}$ are real parameters.

We introduce the notation $\omega_{ik} = \omega_i - \omega_k$ for the energy difference between levels $i$ and $k$. We assume that the one-photon detunings $\omega_{ja,jb} - \omega_0$ are large compared to the Rabi frequencies $\mu_{ja,jb}E_0$ as well as to the two-photon detuning $\delta$. Here, $\omega_0$ and $E_0$ are the characteristic values of the input frequency and input electric field, respectively. In this case, the medium can be considered as an effective two-level system, with two levels $a$ and $b$, evolving slowly in time as compared to the field. To consider the situation where the field may be a short pulse, we express $E$ as a Fourier integral

$$E = \frac{1}{2}\int_{-\infty}^{\infty} e^{-i\omega\tau} E_\omega \, d\omega, \tag{4}$$

with $E_{-\omega} = E_\omega^*$, that is, we take into account all spectral components of the field.

### A. Evolution of the medium state

We use the local coordinates $z$ and $\tau = t - z/c$. As shown in Appendix A, under the far-off-resonance condition, the density matrix of the medium state is governed by the equations

$$\begin{aligned}
\frac{\partial \rho_{aa}}{\partial \tau} &= i(\Omega_{ab}\rho_{ba} - \Omega_{ba}\rho_{ab}) + \gamma_1 \rho_{bb}, \\
\frac{\partial \rho_{bb}}{\partial \tau} &= -i(\Omega_{ab}\rho_{ba} - \Omega_{ba}\rho_{ab}) - \gamma_1 \rho_{bb}, \\
\frac{\partial \rho_{ab}}{\partial \tau} &= i(\Omega_{aa} - \Omega_{bb} + \delta + i\gamma_2)\rho_{ab} + i\Omega_{ab}(\rho_{bb} - \rho_{aa}).
\end{aligned} \tag{5}$$

Here the Stark shifts $\Omega_{aa}$ and $\Omega_{bb}$ and the complex two-photon Rabi frequencies $\Omega_{ab}$ and $\Omega_{ba}$ are given by

$$\Omega_{aa} = \frac{1}{8\hbar}\int_{-\infty}^{\infty}\int_{-\infty}^{\infty} d\omega\, d\omega'\, \alpha_{aa}^{(\omega)} E_\omega E_{\omega'}^* e^{-i(\omega-\omega')\tau},$$

$$\Omega_{bb} = \frac{1}{8\hbar}\int_{-\infty}^{\infty}\int_{-\infty}^{\infty} d\omega\, d\omega'\, \alpha_{bb}^{(\omega)} E_\omega E_{\omega'}^* e^{-i(\omega-\omega')\tau},$$

$$\Omega_{ab} = \frac{1}{8\hbar}\int_{-\infty}^{\infty}\int_{-\infty}^{\infty} d\omega\, d\omega'\, \alpha_{ab}^{(\omega)} E_\omega E_{\omega'+\omega_m}^* e^{-i(\omega-\omega')\tau},$$

$$\Omega_{ba} = \frac{1}{8\hbar}\int_{-\infty}^{\infty}\int_{-\infty}^{\infty} d\omega\, d\omega'\, \alpha_{ba}^{(\omega)} E_\omega E_{\omega'-\omega_m}^* e^{-i(\omega-\omega')\tau}, \tag{6}$$

where $\omega_m = \omega_{ba} - \delta$ is the modulation frequency. The coefficients $\alpha_{ik}^{(\omega)}$ are the Raman polarizability matrix and are given by

$$\begin{aligned}
\alpha_{aa}^{(\omega)} &= \frac{2}{\hbar}\sum_j \frac{\mu_{ja}^2}{\omega_{ja} - \omega}, \\
\alpha_{bb}^{(\omega)} &= \frac{2}{\hbar}\sum_j \frac{\mu_{jb}^2}{\omega_{jb} - \omega}, \\
\alpha_{ab}^{(\omega)} &= \frac{2}{\hbar}\sum_j \frac{\mu_{aj}\mu_{jb}}{\omega_{jb} - \omega}, \\
\alpha_{ba}^{(\omega)} &= \frac{2}{\hbar}\sum_j \frac{\mu_{bj}\mu_{ja}}{\omega_{ja} - \omega}.
\end{aligned} \tag{7}$$

In Eqs. (5) we have added the phenomenological terms which describe the population decay at a rate $\gamma_1$ and the coherence decay at a rate $\gamma_2$.

Note that Eqs. (5) for the medium state have been derived previously for the case of discrete Raman sidebands [13, 29]. In the present case, the field spectrum is continuous and the slowly varying envelope approximation for the time dependence of the field is not used. Therefore, the Stark shifts and the two-photon Rabi frequencies are given in the form of integrals over the continuous variable $\omega$.

### B. Propagation equation for the field in the frequency domain

We study the propagation of the field in the situation where the change of the medium state is negligible. We make the slowly varying envelope approximation for the spatial dependence of the field (the paraxial approximation), that is, we assume that the variation of $E$ with $z$ at constant $\tau$ occurs only over distances much larger than an optical wavelength. In this case, the propagation of the field is, as shown in Appendix B, governed by the equation [13, 29]

$$\frac{\partial E_\omega}{\partial z} = i(\alpha_\omega \rho_{aa} + \beta_\omega \rho_{bb})E_\omega + ig_\omega \rho_{ba} E_{\omega-\omega_m} + ih_\omega \rho_{ab} E_{\omega+\omega_m}. \quad (8)$$

The propagation coefficients $\alpha_\omega$ and $\beta_\omega$ and the coupling coefficients $g_\omega$ and $h_\omega$ are given by

$$\begin{aligned}
\alpha_\omega &= \frac{N\hbar}{\epsilon_0 c}\omega a_\omega, \\
\beta_\omega &= \frac{N\hbar}{\epsilon_0 c}\omega b_\omega, \\
g_\omega &= \frac{N\hbar}{\epsilon_0 c}\omega d_{\omega-\omega_m}, \\
h_\omega &= \frac{N\hbar}{\epsilon_0 c}\omega d_\omega,
\end{aligned} \quad (9)$$

where $N$ is the molecular density. Here we have introduced the notation

$$\begin{aligned}
a_\omega &= \frac{1}{4\hbar}[\alpha_{aa}^{(\omega)} + \alpha_{aa}^{(-\omega)}], \\
b_\omega &= \frac{1}{4\hbar}[\alpha_{bb}^{(\omega)} + \alpha_{bb}^{(-\omega)}], \\
d_\omega &= \frac{1}{4\hbar}[\alpha_{ab}^{(\omega)} + \alpha_{ba}^{(-\omega)}],
\end{aligned} \quad (10)$$

or, equivalently,

$$\begin{aligned}
a_\omega &= \frac{1}{2\hbar^2}\sum_j \left(\frac{\mu_{ja}^2}{\omega_j - \omega_a - \omega} + \frac{\mu_{ja}^2}{\omega_j - \omega_a + \omega}\right), \\
b_\omega &= \frac{1}{2\hbar^2}\sum_j \left(\frac{\mu_{jb}^2}{\omega_j - \omega_b - \omega} + \frac{\mu_{jb}^2}{\omega_j - \omega_b + \omega}\right), \\
d_\omega &= \frac{1}{2\hbar^2}\sum_j \left(\frac{\mu_{aj}\mu_{jb}}{\omega_j - \omega_b - \omega} + \frac{\mu_{aj}\mu_{jb}}{\omega_j - \omega_a + \omega}\right). \quad (11)
\end{aligned}$$

Equation (8) is written in the frequency domain. When the continuous frequency $\omega$ is replaced by discrete Raman sidebands $\omega_q$, Eq. (8) reduces to the propagation equation of Refs. [13, 29]. Equation (8) includes the dispersion in the vicinity of each individual sideband as well as the dispersion from a sideband to another sideband in a multi-sideband spectrum. Note that, in the case where the medium state varies slowly in time, Eq. (8) can be used in conjunction with Eqs. (5), but the frequency component $E_\omega$ becomes time-dependent and, therefore, corresponds to a transient, slowly varying spectrum of the field $E$.

### C. Propagation equation for the Raman sidebands

We consider the case where the spectrum of the field is concentrated in narrow vicinities of the sideband frequencies $\omega_q = \omega_0 + q\omega_m$. Here $q$ is an integer number and $\omega_q > 0$. We represent the electric field $E$ in the form

$$E = \frac{1}{2}\sum_q \left(E_q e^{-i\omega_q \tau} + E_q^* e^{i\omega_q \tau}\right), \quad (12)$$

where $E_q$ is the envelope of the $q$th sideband and is defined by

$$E_q = \int_{\omega_q - \omega_m/2}^{\omega_q + \omega_m/2} E_\omega e^{-i(\omega - \omega_q)\tau} d\omega. \quad (13)$$

For the sideband with the lowest frequency, the lower limit of the integral on the right-hand side of Eq. (13) should be replaced by zero.

For convenience we use the notation

$$f_q = f(\omega_q), \quad f'_q = \left.\frac{df(\omega)}{d\omega}\right|_{\omega_q}, \quad f''_q = \left.\frac{d^2 f(\omega)}{d\omega^2}\right|_{\omega_q}, \quad (14)$$

where $f(\omega)$ can be any of the coefficients $\alpha_\omega$, $\beta_\omega$, $g_\omega$, $h_\omega$, $a_\omega$, $b_\omega$, and $d_\omega$. We expand the coefficients $\alpha_\omega$, $\beta_\omega$, $g_\omega$, and $h_\omega$ in the vicinity of $\omega_q$ to the second order of $\omega - \omega_q$ with the help of the Taylor formula

$$f(\omega) = f_q + f'_q(\omega - \omega_q) + \frac{1}{2}f''_q(\omega - \omega_q)^2. \quad (15)$$

We submit these expansions into the frequency-domain propagation equation (8) and perform the integration from $\omega_q - \omega_m/2$ to $\omega_q + \omega_m/2$. Then, we obtain the following propagation equation for the sideband field envelopes:

$$\begin{aligned}
\frac{\partial E_q}{\partial z} =\ & i[(\alpha_q \rho_{aa} + \beta_q \rho_{bb})E_q + g_q \rho_{ba} E_{q-1} + h_q \rho_{ab} E_{q+1}] \\
& - \left[(\alpha'_q \rho_{aa} + \beta'_q \rho_{bb})\frac{\partial E_q}{\partial \tau} \right.\\
& \left. + g'_q \rho_{ba}\frac{\partial E_{q-1}}{\partial \tau} + h'_q \rho_{ab}\frac{\partial E_{q+1}}{\partial \tau}\right] \\
& - \frac{i}{2}\left[(\alpha''_q \rho_{aa} + \beta''_q \rho_{bb})\frac{\partial^2 E_q}{\partial \tau^2} \right.\\
& \left. + g''_q \rho_{ba}\frac{\partial^2 E_{q-1}}{\partial \tau^2} + h''_q \rho_{ab}\frac{\partial^2 E_{q+1}}{\partial \tau^2}\right]. \quad (16)
\end{aligned}$$

In Eq. (16), the terms proportional to the sideband amplitudes describe the linear propagation of the individual sidebands and the Raman coupling between them. The terms containing the first-order time derivatives account for the effect of the group velocities of the individual sidebands. The terms containing the second-order time derivatives account for the effect of the group velocity dispersion of the individual sidebands. Thus, Eq. (16) is beyond the standard slowly varying envelope approximation, which neglects the time derivatives of the sideband amplitudes. When we apply this approximation, the sideband propagation equation (16) reduces to the slowly varying envelope equation

$$\frac{\partial E_q}{\partial z} = i[(\alpha_q \rho_{aa} + \beta_q \rho_{bb})E_q + g_q \rho_{ba} E_{q-1} + h_q \rho_{ab} E_{q+1}], \quad (17)$$

which was studied in Refs. [13, 29].

### D. Propagation equation for the field in the time domain

We write the electric field $E$ in the form

$$E = \frac{1}{2}(\mathcal{E} + \mathcal{E}^*), \quad (18)$$

where the positive-frequency field component $\mathcal{E}$ is defined by

$$\mathcal{E} = \int_0^{+\infty} E_\omega e^{-i\omega\tau} d\omega = \sum_q E_q e^{-i\omega_q \tau}. \quad (19)$$

We assume that the propagation coefficients $\alpha_\omega$ and $\beta_\omega$ and the coupling coefficients $g_\omega$ and $h_\omega$ vary slowly with $\omega$. We use Eq. (15) to expand $\alpha_\omega$ and $\beta_\omega$ around the characteristic input frequency $\omega_0$. Similarly, we expand $g_\omega$ and $h_\omega$ around the anti-Stokes frequency $\omega_1$ and the Stokes frequency $\omega_{-1}$, respectively. We submit these expansions into the frequency-domain propagation equation (8) and perform the integration from 0 to $+\infty$. Then, with the assumption that $E_\omega = 0$ for $|\omega| < \omega_m$, we obtain

$$\begin{aligned}\frac{\partial \mathcal{E}}{\partial z} &= i[A\rho_{aa} + B\rho_{bb} + F(\tau)]\mathcal{E} \\ &\quad - [A_1 \rho_{aa} + B_1 \rho_{bb} + F_1(\tau)]\frac{\partial \mathcal{E}}{\partial \tau} \\ &\quad - \frac{i}{2}[A_2 \rho_{aa} + B_2 \rho_{bb} + F_2(\tau)]\frac{\partial^2 \mathcal{E}}{\partial \tau^2},\end{aligned} \quad (20)$$

where

$$\begin{aligned}F(\tau) &= K\rho_{ba} e^{-i\omega_m \tau} + Q\rho_{ab} e^{i\omega_m \tau}, \\ F_1(\tau) &= K_1 \rho_{ba} e^{-i\omega_m \tau} + Q_1 \rho_{ab} e^{i\omega_m \tau}, \\ F_2(\tau) &= K_2 \rho_{ba} e^{-i\omega_m \tau} + Q_2 \rho_{ab} e^{i\omega_m \tau}.\end{aligned} \quad (21)$$

Here we have introduced the notation

$$\begin{aligned}A &= \alpha_0 - \omega_0 \alpha_0' + \frac{1}{2}\omega_0^2 \alpha_0'', \\ B &= \beta_0 - \omega_0 \beta_0' + \frac{1}{2}\omega_0^2 \beta_0'', \\ K &= g_1 - \omega_0 g_1' + \frac{1}{2}\omega_0^2 g_1'', \\ Q &= h_{-1} - \omega_0 h_{-1}' + \frac{1}{2}\omega_0^2 h_{-1}'',\end{aligned} \quad (22)$$

$$\begin{aligned}A_1 &= \alpha_0' - \omega_0 \alpha_0'', \\ B_1 &= \beta_0' - \omega_0 \beta_0'', \\ K_1 &= g_1' - \omega_0 g_1'', \\ Q_1 &= h_{-1}' - \omega_0 h_{-1}'',\end{aligned} \quad (23)$$

and

$$\begin{aligned}A_2 &= \alpha_0'', \\ B_2 &= \beta_0'', \\ K_2 &= g_1'', \\ Q_2 &= h_{-1}''.\end{aligned} \quad (24)$$

In Eq. (20), the term containing $\mathcal{E}$ describes the phase shift and the amplification or absorption. The term containing the first-order time derivative describes the effect of the time-varying group velocity. The term containing the second-order time derivative accounts for the effect of the time-varying group velocity dispersion. Note that the time-domain propagation equation (20) for the field amplitude $\mathcal{E}$ can also be derived from Eq. (16) as well as Eq. (17) for the field sidebands $E_q$ if we expand the propagation coefficients $\alpha_q$ and $\beta_q$ around $\omega_0$ and expand the coupling coefficients $g_q$ and $h_q$ around $\omega_1$ and $\omega_{-1}$, respectively.

We now use Eqs. (9) to rewrite Eqs. (22), (23), and (24) in terms of the dispersion coefficients $a_0$ and $b_0$ and the coupling coefficients $d_0$ and $d_{-1}$. The results are

$$\begin{aligned}A &= \frac{N\hbar}{2\epsilon_0 c}\omega_0^3 a_0'', \\ B &= \frac{N\hbar}{2\epsilon_0 c}\omega_0^3 b_0'', \\ K &= \frac{N\hbar}{\epsilon_0 c}\left[\omega_m d_0 - \omega_0 \omega_m d_0' + \frac{\omega_0^2(\omega_0 + \omega_m)}{2}d_0''\right], \\ Q &= \frac{N\hbar}{\epsilon_0 c}\left[-\omega_m d_{-1} + \omega_0 \omega_m d_{-1}' + \frac{\omega_0^2(\omega_0 - \omega_m)}{2}d_{-1}''\right],\end{aligned} \quad (25)$$

$$A_1 = \frac{N\hbar}{\epsilon_0 c}(a_0 - \omega_0 a'_0 - \omega_0^2 a''_0),$$
$$B_1 = \frac{N\hbar}{\epsilon_0 c}(b_0 - \omega_0 b'_0 - \omega_0^2 b''_0),$$
$$K_1 = \frac{N\hbar}{\epsilon_0 c}[d_0 - (\omega_0 - \omega_m)d'_0 - \omega_0(\omega_0 + \omega_m)d''_0],$$
$$Q_1 = \frac{N\hbar}{\epsilon_0 c}[d_{-1} - (\omega_0 + \omega_m)d'_{-1} - \omega_0(\omega_0 - \omega_m)d''_{-1}], \quad (26)$$

and

$$A_2 = \frac{N\hbar}{\epsilon_0 c}(2a'_0 + \omega_0 a''_0),$$
$$B_2 = \frac{N\hbar}{\epsilon_0 c}(2b'_0 + \omega_0 b''_0),$$
$$K_2 = \frac{N\hbar}{\epsilon_0 c}[2d'_0 + (\omega_0 + \omega_m)d''_0],$$
$$Q_2 = \frac{N\hbar}{\epsilon_0 c}[2d'_{-1} + (\omega_0 - \omega_m)d''_{-1}]. \quad (27)$$

Since the medium is far from resonance with the field, we may assume that $a_q \gg \omega_q a'_q \gg \omega_q^2 a''_q$, $b_q \gg \omega_q b'_q \gg \omega_q^2 b''_q$, and $d_q \gg \omega_q d'_q \gg \omega_q^2 d''_q$. With this assumption, we can keep only the dominant terms on the right-hand side of Eqs. (25), (26), and (27). Then, we obtain

$$A = \frac{N\hbar}{2\epsilon_0 c}\omega_0^3 a''_0,$$
$$B = \frac{N\hbar}{2\epsilon_0 c}\omega_0^3 b''_0,$$
$$K = -Q = \frac{N\hbar}{\epsilon_0 c}\omega_m d_0, \quad (28)$$

$$A_1 = \frac{N\hbar}{\epsilon_0 c}a_0,$$
$$B_1 = \frac{N\hbar}{\epsilon_0 c}b_0,$$
$$K_1 = Q_1 = \frac{N\hbar}{\epsilon_0 c}d_0, \quad (29)$$

and

$$A_2 = 2\frac{N\hbar}{\epsilon_0 c}a'_0,$$
$$B_2 = 2\frac{N\hbar}{\epsilon_0 c}b'_0,$$
$$K_2 = Q_2 = 2\frac{N\hbar}{\epsilon_0 c}d'_0. \quad (30)$$

Hence, Eq. (20) reduces to

$$\begin{aligned}\frac{\partial \mathcal{E}}{\partial z} &= i\frac{N\hbar}{\epsilon_0 c}\Big[\frac{1}{2}\omega_0^3(a''_0 \rho_{aa} + b''_0 \rho_{bb}) \\ &\quad + \omega_m d_0(\rho_{ba}e^{-i\omega_m \tau} - \rho_{ab}e^{i\omega_m \tau})\Big]\mathcal{E} \\ &\quad - \frac{N\hbar}{\epsilon_0 c}(a_0 \rho_{aa} + b_0 \rho_{bb} \\ &\quad + d_0 \rho_{ba}e^{-i\omega_m \tau} + d_0 \rho_{ab}e^{i\omega_m \tau})\frac{\partial \mathcal{E}}{\partial \tau} \\ &\quad - i\frac{N\hbar}{\epsilon_0 c}(a'_0 \rho_{aa} + b'_0 \rho_{bb} \\ &\quad + d'_0 \rho_{ba}e^{-i\omega_m \tau} + d'_0 \rho_{ab}e^{i\omega_m \tau})\frac{\partial^2 \mathcal{E}}{\partial \tau^2}. \quad (31)\end{aligned}$$

Similar to the general time-domain propagation equation (20), the simplified off-resonant propagation equation (31) also includes the effect of the dispersion on the scale of multiple sidebands. The coefficient of the term containing $\mathcal{E}$ describes the phase shift per length and the amplification or absorption rate. The coefficients of the terms containing $\partial \mathcal{E}/\partial \tau$ and $\partial^2 \mathcal{E}/\partial \tau^2$ characterize the time-varying group velocity and the time-varying group velocity dispersion, respectively. All these coefficients are modulated by the Raman coherence, that is, by the molecular oscillations.

## III. MODULATION AND COMPRESSION OF A PROBE FIELD BEATING WITH A PREPARED RAMAN COHERENCE

### A. Model and propagation equation in the case of zero dispersion

We assume that the Raman coherence $\rho_{ab}$ is established adiabatically by using two long off-resonant driving fields [13, 29]. After the Raman coherence is prepared, we send a weak probe field $E$ into the medium. When all dispersion effects, except for the group velocity and the group velocity dispersion, are negligible, the propagation of the probe field is governed by Eq. (20).

We consider the case where the dispersion of the medium is negligible, that is, $a_q = a_0$, $b_q = b_0$, $d_q = d_0$, and, consequently, $a'_q = b'_q = d'_q = a''_q = b''_q = d''_q = 0$. In this case, Eqs. (25), (26), and (27) yield $A = B = A_2 = B_2 = K_2 = Q_2 = 0$, $K = -Q = (N\hbar/\epsilon_0 c)\omega_m d_0$, $K_1 = Q_1 = (N\hbar/\epsilon_0 c)d_0$, $A_1 = (N\hbar/\epsilon_0 c)a_0$, and $B_1 = (N\hbar/\epsilon_0 c)b_0$. Hence, Eq. (20) for the positive-frequency component $\mathcal{E}$ leads the following equation for the probe field $E$ [16, 19]:

$$\frac{\partial E}{\partial z} = -\frac{N\hbar}{\epsilon_0 c}\frac{\partial}{\partial \tau}(a_0 \rho_{aa} + b_0 \rho_{bb} + d_0 \rho_{ba}e^{-i\omega_m \tau} + d_0 \rho_{ab}e^{i\omega_m \tau})E. \quad (32)$$

When we compare Eq. (32) with the reduced wave equa-

tion

$$\frac{\partial E}{\partial z} = -\frac{1}{2\epsilon_0 c}\frac{\partial P}{\partial \tau}, \quad (33)$$

we find that the instantaneous susceptibility $\chi = P/\epsilon_0 E$ of the polarization $P$ to the probe field $E$ is given by

$$\chi = \frac{2N\hbar}{\epsilon_0}(a_0\rho_{aa} + b_0\rho_{bb} + d_0\rho_{ba}e^{-i\omega_m\tau} + d_0\rho_{ab}e^{i\omega_m\tau}). \quad (34)$$

Note that this susceptibility is modulated in time by the prepared Raman coherence $\rho_{ab}$ at the Raman frequency $\omega_m$.

Using the definition (34) for the time-varying instantaneous susceptibility $\chi$, we can rewrite the propagation equation (32) in the form

$$\frac{\partial E}{\partial z} = -\frac{1}{2c}\frac{\partial \chi}{\partial \tau}E - \frac{1}{2c}\chi\frac{\partial E}{\partial \tau}. \quad (35)$$

The coefficient $(1/2c)\partial\chi/\partial\tau$ of the first term on the right-hand side is proportional to the slope of the time-varying susceptibility $\chi$. This coefficient determines gain or absorption, depending on its sign. Meanwhile, the coefficient $\chi/2c$ of the second term on the right-hand side is the reciprocal of the group velocity in the local time. Because of the absence of dispersion, the group velocity is equal to the phase velocity.

In the coherence preparation step, with the assumptions of large two-photon detuning, limited bandwidth, and negligible dispersion, the medium follows an adiabatic eigenstate with [13, 29]

$$\rho_{aa} = \cos^2\theta, \qquad \rho_{bb} = \sin^2\theta, \quad (36)$$

and

$$\rho_{ab} = e^{i(\phi_0 - \kappa z)}\sin\theta\cos\theta. \quad (37)$$

Here $\phi_0$ is the relative phase between the driving fields at the input, $\kappa$ is the phase shift per length of the prepared Raman coherence and is given by

$$\kappa = \frac{N\hbar}{\epsilon_0 c}\omega_m(a_0\rho_{aa} + b_0\rho_{bb}), \quad (38)$$

and $\theta$ is determined by the equation

$$\tan 2\theta = \left|\frac{2\Omega_{ab}^{(\text{ext})}}{\delta + \Omega_{aa}^{(\text{ext})} - \Omega_{bb}^{(\text{ext})}}\right|\text{sgn}(\delta), \quad (39)$$

where $|\theta| \leq \pi/2$. In Eq. (39), $\Omega_{aa}^{(\text{ext})}$ and $\Omega_{bb}^{(\text{ext})}$ are the Stark shifts and $\Omega_{ab}^{(\text{ext})}$ is the two-photon Rabi frequency, produced by the driving fields. We assume that the probe field is weak and short as compared to the driving fields so that the medium state is mainly determined by the driving fields and is not substantially affected by the probe field.

## B. Analytical solution and conservation relations

We now use the reduced local time $\eta = \tau - z/v + \phi/\omega_m$, where $v = \omega_m/\kappa$ and $\phi = \phi_0 + (\pi/2)\text{sgn}(\delta)$. Then, the propagation equation (32) simplifies to

$$\frac{\partial E}{\partial z} = -\frac{\alpha}{\omega_m}\frac{\partial}{\partial \eta}\sin(\omega_m\eta)E. \quad (40)$$

Here we have introduced the coupling parameter

$$\alpha = \frac{2\hbar}{\epsilon_0 c}N\omega_m d_0\rho_0, \quad (41)$$

where $\rho_0 = |\sin\theta\cos\theta| = |\rho_{ab}|$. The solution of Eq. (40) is [16]

$$E(z,\eta) = E_{\text{in}}(s)G(\eta), \quad (42)$$

where $E_{\text{in}}(s) = E(z = 0, s)$ is the input field, the input time $s$ is determined from the output time $\eta$ by the equation

$$\tan(\omega_m s/2) = e^{-\alpha z}\tan(\omega_m\eta/2), \quad (43)$$

and the function $G(\eta) = \sin(\omega_m s)/\sin(\omega_m\eta)$ is given by [19]

$$G(\eta) = \frac{1}{e^{\alpha z}\cos^2\frac{\omega_m\eta}{2} + e^{-\alpha z}\sin^2\frac{\omega_m\eta}{2}}. \quad (44)$$

It follows from Eq. (43) that

$$ds/d\eta = G(\eta). \quad (45)$$

When the output time $\eta$ is close to a given time $\eta_i$, we find from Eq. (45) that the input time $s$ can be approximated as

$$s(\eta) = s(\eta_i) + G(\eta_i)(\eta - \eta_i). \quad (46)$$

Hence, for the input frequency $\omega_0$, the instantaneous frequency $\omega_{\text{osc}}(\eta_i)$ of the optical oscillations of $E$ in the vicinity of the given time $\eta_i$ is $G(\eta_i)\omega_0$, that is,

$$\omega_{\text{osc}}(\eta) = G(\eta)\omega_0 = \frac{\omega_0}{e^{\alpha z}\cos^2\frac{\omega_m\eta}{2} + e^{-\alpha z}\sin^2\frac{\omega_m\eta}{2}}. \quad (47)$$

Equation (47) shows that the instantaneous oscillation frequency $\omega_{\text{osc}}(\eta)$ is modulated in time.

According to Eqs. (42) and (45), the height and duration of optical oscillations in the probe field are changed by the reciprocal factors $E(z,\eta)/E_{\text{in}}(s) = G(\eta)$ and $d\eta/ds = 1/G(\eta)$, respectively. A consequence of these equations is the relation

$$\int_{\eta_1}^{\eta_2}E(z,\eta)\,d\eta = \int_{s_1}^{s_2}E_{\text{in}}(s)\,ds, \quad (48)$$

which means the conservation of the pulse area in any given time interval $(\eta_1, \eta_2)$.

Using Eqs. (42), (45), and (47), we can show that

$$\frac{c\epsilon_0}{2}\int_{\eta_1}^{\eta_2}\frac{E^2(z,\eta)}{\hbar\omega_{\rm osc}(\eta)}\,d\eta=\frac{c\epsilon_0}{2}\int_{s_1}^{s_2}\frac{E_{\rm in}^2(s)}{\hbar\omega_0}\,ds. \quad (49)$$

The expression on the right-hand side of the above equation is the number of photons in the input field, with the frequency $\omega_0$. The expression on the left-hand side is the number of photons in the output field, with the instantaneous frequency $\omega_{\rm osc}(\eta)$. Therefore, Eq. (49) describes the conservation of the photon number. This is a very general relation which is always satisfied for Raman processes [12, 13].

It follows from Eqs. (45) and (47) that

$$s_2-s_1=\int_{\eta_1}^{\eta_2}G(\eta)\,d\eta=\int_{\eta_1}^{\eta_2}\frac{\omega_{\rm osc}(\eta)}{\omega_0}\,d\eta. \quad (50)$$

We define the mean frequency of oscillations in the time interval $(\eta_1,\eta_2)$ by

$$\bar{\omega}(\eta_1,\eta_2)=\frac{1}{\eta_2-\eta_1}\int_{\eta_1}^{\eta_2}\omega_{\rm osc}(\eta)\,d\eta. \quad (51)$$

With this definition, Eq. (50) yields

$$\bar{\omega}(\eta_1,\eta_2)(\eta_2-\eta_1)=\omega_0(s_2-s_1). \quad (52)$$

Equation (52) indicates that the product of the pulse length and the mean frequency is constant during the propagation process. Furthermore, we have $E(z,\eta)=0$ if $E_{\rm in}(s)=0$ and vice versa, that is, a zero of the input field at an input time $s$ leads to a zero of the output field at the corresponding output time $\eta$. Hence, the number of optical oscillations is conserved. This conservation is a simple consequence of the pulse reshaping described by Eq. (42).

The above conservation relations are especially useful for the analysis in the case where the pulse duration is much shorter than the molecular period. While the photon number conservation is very general, the other conservation relations obtained by us are approximate and are seriously violated when dispersion or nonadiabatic effects are substantial.

To get deeper insight into the solution (42), we rewrite the reduced propagation equation (40) as

$$\frac{\partial E}{\partial z}=-\alpha\cos(\omega_m\eta)E-\frac{\alpha}{\omega_m}\sin(\omega_m\eta)\frac{\partial E}{\partial\eta}. \quad (53)$$

The coefficient $\alpha\cos(\omega_m\eta)$ of the first term on the right-hand side is a time-varying gain or absorption rate, depending on its sign. The coefficient $(\alpha/\omega_m)\sin(\omega_m\eta)$ of the second term on the right-hand side is a measure of the time-varying group and phase velocities. The modulation in the amplitude of the solution (42) is due to the first coefficient, and the modulation in the instantaneous oscillation frequency is due to the second coefficient. Both coefficients originate from the time-varying susceptibility $\chi$, given by Eq. (34), which in terms of the reduced local time $\eta$ takes the form

$$\chi=\frac{2c}{\omega_m}[\kappa+\alpha\sin(\omega_m\eta)]. \quad (54)$$

The gain or absorption rate $\alpha\cos(\omega_m\eta)$ is determined by the slope of the inversed time-varying group velocity $(\alpha/\omega_m)\sin(\omega_m\eta)$, or more exactly, by the slope of the time-varying susceptibility $\chi$. This gain (absorption) is accompanied by the increase (decrease) in the oscillation frequency. Since the photon number is conserved [Eq. (49)], this gain (absorption) is not the conventional gain (absorption), in which the photon number increases (decreases).

In the particular case where $\alpha z\ll 1$ and $\omega_m\ll\omega_0$, we can neglect the first term on the right-hand side of Eq. (53). In this case, for the initial condition $E_{\rm in}(s)=\mathcal{E}_0^{({\rm in})}\cos(\omega_0 s)$, we find the approximate solution $E(z,\eta)=\mathcal{E}_0^{({\rm in})}\cos[\omega_0\int_0^\eta G(x)\,dx]$, which shows frequency chirping but no compression or stretching.

### C. Frequency conversion and pulse compression due to the time-varying susceptibility

For any fixed value of $z$, the function $G$ has peaks and dips at equidistant times

$$\eta_n=(2n+1)\pi/\omega_m \quad (55)$$

and

$$\eta_{n-1/2}=2n\pi/\omega_m, \quad (56)$$

respectively. For $\eta=\eta_n$, we have $\cos(\omega_m\eta/2)=0$ leading to $G=e^{\alpha z}>1$. This value indicates a pulse compression in the vicinity of $\eta_n$, with the compression factor $e^{\alpha z}$ [16]. For $\eta=\eta_{n-1/2}$, we have $\sin(\omega_m\eta/2)=0$ leading to $G=e^{-\alpha z}<1$. This value indicates a pulse stretching in the vicinity of $\eta_{n-1/2}$. Compression and stretching, occurring sequentially at $\eta_n$ and $\eta_{n-1/2}$, lead to the formation of a train of short pulses with separation $T_m=2\pi/\omega_m$ (the Raman period) and pulse length $T_m e^{-\alpha z}$. The peak frequency and peak amplitude of the pulses in the train grow with $e^{\alpha z}$. Such pulse compression and stretching are due to the modulation of the gain-or-absorption coefficient $\alpha\cos(\omega_m\eta)$, which is the slope of the time-varying susceptibility $\chi$ [see Eq. (54)]. Thus, the time-varying susceptibility of the medium acts like a temporal grating which compresses and decompresses the probe at certain times.

Note that the generation of an almost periodic train of subfemtosecond pulses in multicascade stimulated Raman scattering has been predicted by Kaplan and Shkolnikov [27]. They have shown that Raman active materials can support solitons consisting of a pump laser wave and many Raman-coupled components. Due to mode locking all these soliton components propagate with the same group velocity, have the same amplitude shape, and fully overlap in time and space. The coherent interference of these mode-locked frequency components gives rise to a train of subfemtosecond pulses. We wish to point out that the work [27] implicitly assumed phase matching among the Raman components. Another method for generation of subfemtosecond solitons is based on the idea of electromagnetic bubbles [30]. According to this method, unipolar, supershort, and intense nonoscillating solitary pulses of electromagnetic radiation can be generated in a gas of two-level or classically nonlinear atoms by half-cycle pulses or short laser pulses. The key difference between the short-solitary-pulse generation schemes [27, 30] and our scheme is that solitons are stationary waves that propagate over substantial distances with unchanged shape and length, while the train of short pulses formed in our scheme is nonstationary, results from the beating of a weak probe with a prepared Raman coherence, and has a compression factor increasing with the propagation distance.

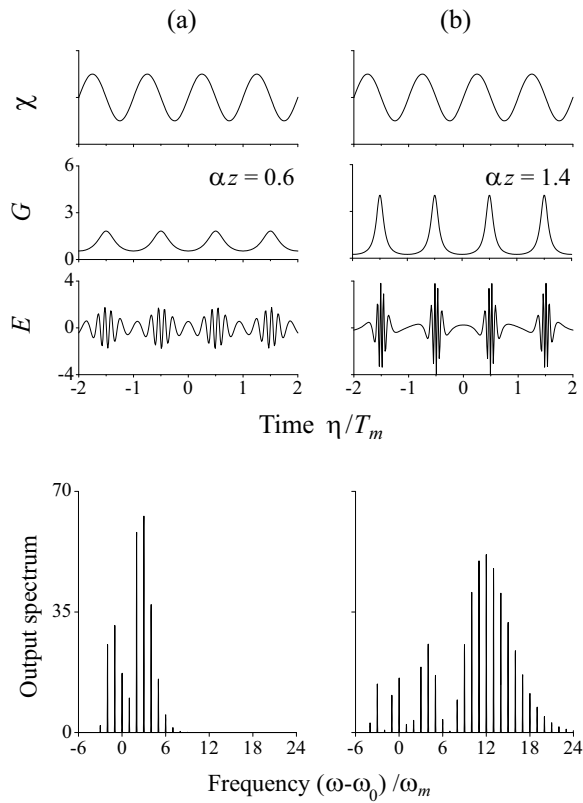

FIG. 2: Compression of a probe pulse, calculated from Eq. (42) for $\alpha z = 0.6$ (a) and 1.4 (b). The carrier frequency, pulse width, and peak time of the input probe field are $\omega_0 = 5.2\,\omega_m$, $T = 10\,T_m$, and $\eta_p = 0$, respectively. The plots for the susceptibility $\chi$ as a function of time show only the shape but not the exact magnitude of this function. The plots for the output field $E$ and for the spectral intensity $|E_\omega|^2$ are normalized to the peak value of the input field.

To see the relation between the modulation of the susceptibility and the formation of a train of compressed pulses, we illustrate in Fig. 2 the compression of a probe pulse, calculated from Eq. (42) for $\alpha z = 0.6$ (a) and 1.4 (b). The carrier frequency, pulse width, and peak time of the input probe field are $\omega_0 = 5.2\,\omega_m$, $T = 10\,T_m$, and $\eta_p = 0$, respectively. As seen from the plots for the susceptibility $\chi$ and the function $G$, the modulation of $\chi$ generates the comb structure of $G$, with peaks (dips) at the times corresponding to the steepest negative (positive) slopes of $\chi$. This leads to the formation of a train of compressed pulses and to the generation of new Raman sidebands (see the plots for the field $E$ and for the spectral intensity $|E_\omega|^2$). We observe periodic changes in the oscillation frequency as the time increases. The spectrum consists of sharp Raman sidebands, most of them are on the anti-Stokes region of frequency. When the factor $\alpha z$ is increased, we observe an increase in the height and a decrease in the width of the pulses in the output train. Furthermore, we observe that the number of generated anti-Stokes sidebands increases quickly with $\alpha z$ while the number of Stokes sidebands does not change substantially.

The periodic change in the optical oscillation frequency of $E$, observed in Fig. 2, is in agreement with Eq. (47) for $\omega_{\rm osc}(\eta)$. In particular, in the vicinity of $\eta_n$ (compression region) or $\eta_{n-1/2}$ (stretching region), the input time $s$ is related to the output time $\eta$ as $s = \eta_n + e^{\alpha z}(\eta - \eta_n)$ or $s = \eta_{n-1/2} + e^{-\alpha z}(\eta - \eta_{n-1/2})$, respectively. And the frequency of the field increases or decreases, respectively, by the factor $e^{\alpha z}$.

When the duration $T$ of the input field $E_{\rm in}$ is long compared to the Raman period $T_m$, the maximal and minimal frequencies of the optical oscillations of $E$ are $\omega_{\max} = e^{\alpha z}\omega_0$ and $\omega_{\min} = e^{-\alpha z}\omega_0$, respectively. The oscillations at these frequencies generate the anti-Stokes and Stokes Raman sidebands with the order numbers

$$q_{AS} = (e^{\alpha z} - 1)\frac{\omega_0}{\omega_m} \qquad (57)$$

and

$$q_S = -(1 - e^{-\alpha z})\frac{\omega_0}{\omega_m}, \qquad (58)$$

respectively. For increasing $\alpha z$, the anti-Stokes sideband order $q_{AS}$ increases as $(e^{\alpha z}-1)$ while the Stokes sideband order $q_S$ changes slowly and tends to the limiting value $-\omega_0/\omega_m$ (see the plots for the spectral intensity in Fig. 2). In the particular case where $\alpha z \ll 1$, we have

$$q_{AS} = -q_S = \frac{\omega_0}{\omega_m}\alpha z = \gamma z, \qquad (59)$$

where

$$\gamma = \frac{\omega_0}{\omega_m}\alpha = \frac{2\hbar}{\epsilon_0 c}N\omega_0 d_0 \rho_0. \qquad (60)$$

This result is in agreement with the estimate for the number of lines generated by two driving fields in a medium with limited modulation bandwidth and zero dispersion [13]. It should be emphasized here that the Raman sideband orders $q_{AS}$ and $q_S$ approximate but do not determine precisely the actual size of the generated Raman spectrum.

When the duration $T$ of the input field $E_{\text{in}}$ is much longer than the Raman period $T_m$, the structure of the generated pulse train is not sensitive to the timing of the input. However, when $T$ is comparable to or smaller than $T_m$, the structure and number of pulses in the output depend on the relative position of the input peak time $\eta_p$ with respect to $\eta_n$ and $\eta_{n-1/2}$. In particular, when $T$ is short compared to $T_m$, the output may be a single compressed or stretched pulse. Indeed, in this case, if $\eta_p$ is in the vicinity of $\eta_n$ ($\eta_{n-1/2}$), then the probe pulse will be compressed (stretched) and the frequency of the optical oscillations will increase (decrease).

To see the sensitivity of the pulse shaping to the injection time of the input, we illustrate in Fig. 3 the compression or stretching of a probe pulse with a length short compared to the Raman period. The plots are calculated from Eq. (42) for the injection time $\eta_p/T_m = 0$ (a), 0.25 (b), 0.5 (c), and 0.75 (d). The carrier frequency and pulse width of the input probe field are $\omega_0 = 15.2\,\omega_m$ and $T = 0.1\,T_m$, respectively. The medium length and the prepared Raman coherence are such that $\alpha z = 0.8$. The dotted lines show $G$ as a function of time. The plots show that, when the input field is sent in at a time corresponding to a dip [Fig. 3(a)] or a peak [Fig. 3(c)] of the function $G$, the output is stretched or compressed, the oscillation frequency decreases or increases, and the generated output spectrum is confined in the region of Stokes or anti-Stokes sidebands, respectively. Note that the output spectrum is continuous because the input pulse is short and, therefore, its Fourier-transform limited width covers several Raman sidebands. When the input is injected at a time between a dip and a peak of $G$ [Figs. 3(b) and 3(d)], the temporal behavior of the output exhibits both compression (in the region near to the peak of $G$) and stretching (in the region near to the dip of $G$), and the output spectrum is spread in both the Stokes and anti-Stokes regions.

### D. Raman spectrum

We calculate the spectrum of the propagating probe field in the case where the input is a monochromatic field, given by

$$E_{\text{in}}(s) = \mathcal{E}_0^{(\text{in})} \cos(\omega_0 s), \quad (61)$$

and the input pulse length $T$ is large compared to the Raman period $T_m$. Using the formula [31] $\sinh x/(\cosh x - \cos t) = 1 + 2\sum_{n=1}^{\infty} e^{-nx}\cos nt$ for $x > 0$, we can expand the function $G(\eta)$ into a Fourier series as

$$G(\eta) = 1 + 2\sum_{n=1}^{\infty}(-1)^n \tanh^n\left(\frac{\alpha z}{2}\right)\cos(n\omega_m \eta). \quad (62)$$

When we substitute this expansion into Eq. (45) for the input time $s$ and integrate the result, we obtain

$$s = \eta + 2\sum_{n=1}^{\infty}\frac{(-1)^n}{n\omega_m}\tanh^n\left(\frac{\alpha z}{2}\right)\sin(n\omega_m \eta). \quad (63)$$

Then, with the use of the generating function [31] $e^{i\xi \sin \theta} = \sum_{k=-\infty}^{\infty} J_k(\xi)e^{ik\theta}$ of the Bessel functions $J_k$, we can expand the complex function $e^{i\omega_0 s}$ in the form

$$e^{i\omega_0 s} = e^{i\omega_0 \eta}\prod_{n=1}^{\infty}\sum_{k=-\infty}^{\infty}(-1)^{kn}J_k\left[\frac{2\omega_0}{n\omega_m}\tanh^n\left(\frac{\alpha z}{2}\right)\right] \\ \times e^{ikn\omega_m \eta}. \quad (64)$$

We use Eq. (64) to expand the expression (61) for the input field $E_{\text{in}}(s)$ into a Fourier series in terms of the output time $\eta$. We substitute the resultant expansion and the expansion (62) into Eq. (42) for the output field $E$. Then, the sideband $\mathcal{E}_q$ of the output field $E = \text{Re}\sum_q \mathcal{E}_q e^{-i\omega_q \eta}$ is found to be

$$\mathcal{E}_q = (-1)^q \mathcal{E}_0^{(\text{in})}\sum_{l=-\infty}^{\infty}\tanh^{|l|}\left(\frac{\alpha z}{2}\right) \\ \times \sum_{\{k_n\}_{q-l}}\prod_{n=1}^{\infty}J_{k_n}\left[\frac{2\omega_0}{n\omega_m}\tanh^n\left(\frac{\alpha z}{2}\right)\right]. \quad (65)$$

Here $\{k_n\}_{q-l}$ is a set of integers $k_n$ satisfying the condition $\sum_{n=1}^{\infty}k_n n = q - l$. Equation (65) shows that, in the case where the input pulse length $T$ is large compared to the Raman period $T_m$, the spectrum of the probe field $E$ is discrete and exhibits oscillations governed by the Bessel functions (see the plots for the spectral intensity in Fig. 2). When $T$ is comparable to or smaller than $T_m$, the spectrum of the envelope $\mathcal{E}_0^{(\text{in})}$ with respect to the output time $\eta$ may cover two or more Raman lines. In this case, the generated Raman sidebands may have overlapping profiles, the discrete structure of the Raman sidebands may be suppressed, and therefore the spectrum of $E$ may become continuous (see the plots for the output spectrum in Fig. 3).

We now show that the Bessel function solution obtained earlier [13] for the Raman sidebands is a particular case of the general Bessel function series (65). For this purpose we consider the situation where $\omega_0 \geq \omega_m$ and

$$\tanh\left(\frac{\alpha z}{2}\right) \ll 1, \quad \frac{\omega_0}{\omega_m}\tanh^2\left(\frac{\alpha z}{2}\right) \ll 1. \quad (66)$$

Then, using the property $J_k(0) = \delta_{k,0}$, we find that the dominant term on the right-hand side of Eq. (65) is the term with $l = k_2 = k_3 = \cdots = 0$ and $k_1 = q$. Hence, we obtain

$$\mathcal{E}_q = (-1)^q \mathcal{E}_0^{(\text{in})}J_q\left[\frac{2\omega_0}{\omega_m}\tanh\left(\frac{\alpha z}{2}\right)\right]. \quad (67)$$

Since $\omega_0 \geq \omega_m$, the conditions (66) lead to

$$\alpha z \ll \sqrt{\frac{\omega_m}{\omega_0}} \leq 1, \quad (68)$$

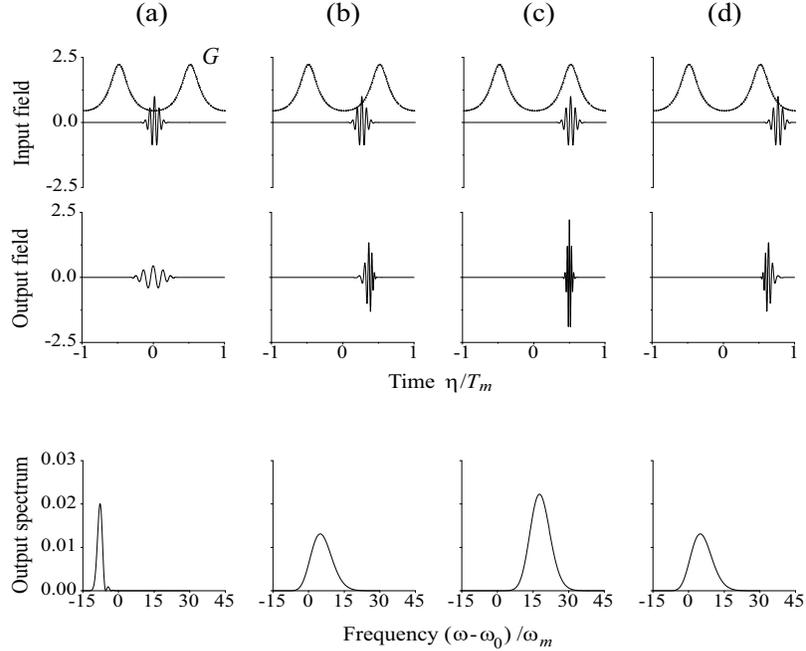

FIG. 3: Compression or stretching of a probe pulse with a length short compared to the Raman period. The plots are calculated from Eq. (42) for the parameter $\alpha z = 0.8$. The carrier frequency and pulse width of the input probe field are $\omega_0 = 15.2\,\omega_m$ and $T = 0.1\,T_m$, respectively. The injection time is $\eta_p/T_m = 0$ (a), 0.25 (b), 0.5 (c), and 0.75 (d). The plots for the input field, output field, and output spectrum are normalized to the peak value of the input field. The dotted lines show $G$ as a function of time. Note that the number of optical oscillations is the same for all time-domain plots.

that is,

$$\gamma z \ll \sqrt{\frac{\omega_0}{\omega_m}}. \qquad (69)$$

The condition (68) yields $e^{\alpha z} \cong 1$, that is, the pulse compression is not strong. The condition (69) and Eq. (59) give $q_{AS} = -q_S \ll (\omega_0/\omega_m)^{1/2}$, that is, the bandwidth of the generated spectrum is limited by the value $(\omega_0/\omega_m)^{1/2}$.

When we use the approximation $\tanh(\alpha z/2) \cong \alpha z/2$, Eq. (67) reduces to the Bessel function solution [13]

$$\mathcal{E}_q = (-1)^q \mathcal{E}_0^{(\text{in})} J_q(\gamma z). \qquad (70)$$

The Fourier synthesis of these Raman sidebands yields the field $E = \mathcal{E}_0^{(\text{in})} \cos[\omega_0 \eta - \gamma z \sin(\omega_m \eta)]$, which is modulated in frequency but is not compressed.

If we extend the above results for the case where the input consists of two fields, at $\omega_0$ and $\omega_{-1}$, then instead of Eq. (70) we will obtain

$$\mathcal{E}_q = (-1)^q \mathcal{E}_0^{(\text{in})} J_q(\gamma z) + (-1)^{q+1} \mathcal{E}_{-1}^{(\text{in})} J_{q+1}(\gamma z). \qquad (71)$$

This expression is in full agreement with the earlier result for the sideband spectrum generated by two driving fields in a medium with limited modulation bandwidth and zero dispersion [13]. The corresponding output field is $E = \mathcal{E}_0^{(\text{in})} \cos[\omega_0 \eta - \gamma z \sin(\omega_m \eta)] + \mathcal{E}_{-1}^{(\text{in})} \cos[\omega_{-1}\eta - \gamma z \sin(\omega_m \eta)]$. Thus, the analytical Bessel function solution [13] is valid when the condition (69) or, equivalently, the condition (68) is satisfied. Under these conditions, the bandwidth of the generated spectrum is limited $[q_{AS} = -q_S \ll (\omega_0/\omega_m)^{1/2}]$ and the pulse compression by the time-varying group velocity is not strong ($e^{\alpha z} \cong 1$).

### E. Effect of group velocity dispersion

In addition to the direct pulse compression by group velocity modulation, there is another mechanism: Pulse chirping, followed by GVD, can produce single-cycle pulses [13]. To study the effect of GVD on pulse compression, we perform numerical calculations for a realistic system, namely, for molecular hydrogen with the fundamental vibrational transition $Q_1(0)$ at frequency $\omega_m = 4149.7$ cm$^{-1}$ and with the solid-state density $N = 2.6 \times 10^{22}$ cm$^{-3}$ [29, 32]. We use this high-density molecular hydrogen as a model of solid hydrogen. The characteristic property of the majority of molecular crystals including solid hydrogen is that the molecules in these solids retain their identity and that their intrinsic properties are modified only slightly by the intermolecular interaction [32]. The advantages of solid hydrogen as a Raman medium are that it has a high number density, short medium length, small dephasing rate, large coherence length, and negligible phase mismatch [8, 33]. In addition, the exper-iment with a solid can be performed in vacuum, without the need for intermediate windows, in order to prevent chirping of the generated subfemtosecond pulses. The use of nanosecond pumping for the narrow vibrational transition $Q_1(0)$ in solid hydrogen allows us to prepare a substantial coherence at low pump intensity, avoiding the counteraction of the Kerr effect and nonlinear dispersion. Furthermore, without the need for external phase compensation, the probe pulse can be compressed to a triplet, doublet, or even singlet of subfemtosecond pulses [19]. Another advantage of solid hydrogen as a Raman medium is that, in a pump-probe experiment for short-pulse generation, one can achieve beam separation at the output by using a small tilt angle between the applied fields [34], which does not affect phase matching or beam overlap in the medium, since the crystal is thin.

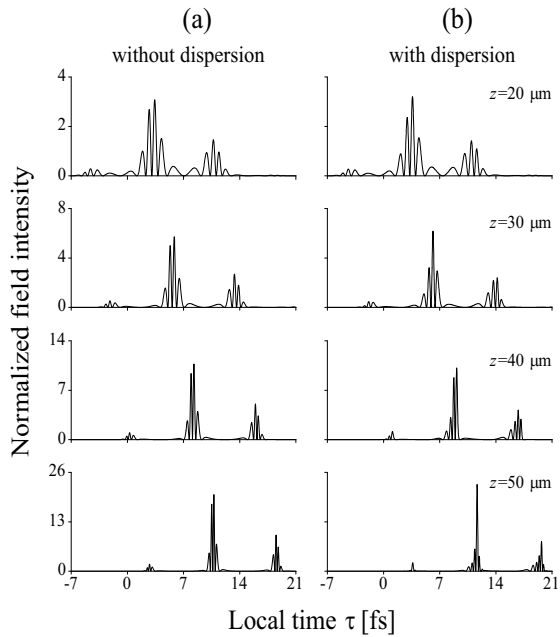

FIG. 4: Comparison of the temporal profiles of a probe field calculated for the parameters of solid hydrogen with zero (a) and nonzero (b) dispersion. The medium lengths are $z = 20$, 30, 40, and 50 $\mu$m (see the labels in the plots on the right-hand side). The medium is prepared in an antiphased state with $\rho_{aa} = 0.85$, $\rho_{bb} = 0.15$, and $|\rho_{ab}| = 0.36$. For the input field we use a pulse with a central wavelength of 800 nm, a pulse length of 10 fs, and a peak time $\tau_p = 0$.

We use the propagation equations (32) and (31) to calculate the temporal profiles of the probe field in the cases of zero and nonzero dispersion, respectively, and plot the results in Figs. 4(a) and 4(b), respectively. For the input probe field we use a Gaussian pulse with a central wavelength of 800 nm ($\omega_0 = 12{,}500$ cm$^{-1}$), a pulse length of 10 fs, and a peak time $\tau_p = 0$. The medium state is calculated from Eqs. (36) and (37) for the parameters $\theta = -0.4$ and $\phi_0 = 0$, which correspond to an antiphased state with $\rho_{aa} = 0.85$, $\rho_{bb} = 0.15$, and $|\rho_{ab}| = 0.36$.

The dispersion and coupling coefficients are calculated from the dipole moments and level energies of parahydrogen [29] and are $a_0 = 2.42 \times 10^{-7}$, $b_0 = 2.63 \times 10^{-7}$, and $d_0 = 5.50 \times 10^{-8}$, in the SI units. The first-order derivatives of the dispersion and coupling coefficients with respect to the frequency are $a'_0 = 3.13 \times 10^{-24}$, $b'_0 = 3.81 \times 10^{-24}$, and $d'_0 = 1.25 \times 10^{-24}$, in the SI units. The second-order derivatives of the dispersion and coupling coefficients with respect to the frequency are $a''_0 = 1.41 \times 10^{-39}$, $b''_0 = 1.73 \times 10^{-39}$, and $d''_0 = 5.07 \times 10^{-40}$, in the SI units. We show the temporal profiles of the field intensity calculated for the medium lengths $z = 20$, 30, 40, and 50 $\mu$m (see the labels on the right-hand side of the figure). These lengths correspond to the values $\alpha z = 0.64$, 0.95, 1.27, and 1.59, respectively, that is, to the compression factors $e^{\alpha z} = 1.89$, 2.60, 3.57, and 4.91, respectively.

Comparison between the plots in Figs. 4(a) and 4(b) shows that the dispersion of the medium does not affect much pulse compression in the cases of $z = 20$, 30, and 40 $\mu$m. In the case of $z = 50$ $\mu$m, the dispersion improves pulse compression by a factor of about 3. Additional numerical calculations for larger medium lengths show that, when the medium length is too large, for example, when $z = 75$ $\mu$m, the dispersion of the medium reduces pulse compression.

The optimal condition for pulse compression by GVD has been derived earlier for the case where the direct compression by group velocity modulation is negligible [13]. According to them, the optimal compression may be obtained when the relative group delay $\Delta\tau_D$ corresponding to the generated spectral bandwidth $\Delta\omega = \omega_{\max} - \omega_{\min}$ is equal to the Raman half-period $T_m/2$. The reason is the following: if dispersion is absent, different frequency components will emerge at different times during a Raman half-period [see Eq. (47) for $\omega_{\mathrm{osc}}(\eta)$ and see also the plots for $E$ in Fig. 2]. However, when GVD is present, the group delays of the frequency components are different from each other. If the difference of the group delays compensates the difference of the dispersionless starting times, all the frequency components of the field will emerge at the same time and will interfere with each other. Then, pulse compression may occur.

We extend this argument to include the effects of both susceptibility modulation and GVD. We assume that the length of the input field is long compared to the Raman period. In this case, the maximal and minimal oscillation frequencies are $\omega_{\max} = e^{\alpha z}\omega_0$ and $\omega_{\min} = e^{-\alpha z}\omega_0$, respectively. These values give the bandwidth $\Delta\omega = 2\omega_0 \sinh(\alpha z)$. The group velocity dispersion, measured by the second-order derivative $k''(\omega)$ of the propagation constant $k(\omega)$, is assumed to be invariant and is estimated by $k''(\omega_0)$. From the coefficients of the terms in the general frequency-domain propagation equation (8) or, similarly, in the time-domain equation (20), we find $k''(\omega_0) = (N\hbar/\epsilon_0 c)[(2a'_0 + \omega_0 a''_0)\rho_{aa} + (2b'_0 + \omega_0 b''_0)\rho_{bb}]$. Keeping only the dominant terms yields $k''(\omega_0) = (2N\hbar/\epsilon_0 c)(a'_0\rho_{aa} + b'_0\rho_{bb})$, in agreement with the coefficients of the terms in the simplified propagation equation (31). Since the group delay of a frequency component is determined by $\tau_D(\omega) = z/v_g(\omega) = zk'(\omega)$, the relative group delay $\Delta\tau_D$ is approximated by $\Delta\tau_D = k''(\omega_0)\Delta\omega\, z$. Hence, it follows from the condition $\Delta\tau_D = T_m/2$ that the medium length $L$ required for the optimal effect of GVD on pulse compression is determined by

$$L\sinh(\alpha L) = \frac{\pi}{2\omega_m\omega_0 k''(\omega_0)}. \tag{72}$$

At this medium length, if the compression due to the destructive interference between the generated sidebands is optimum, the compression factor $\Gamma$, defined as the ratio of pulse separation to pulse width, is estimated to be equal to the number of sidebands [13]. We approximate this number by $q_{\text{total}} = q_{AS} - q_S + 1$, where the sideband orders $q_{AS}$ and $q_S$ are given by Eqs. (57) and (58), respectively. Then, we have

$$\Gamma = 1 + 2(\omega_0/\omega_m)\sinh(\alpha L). \tag{73}$$

This factor corresponds to the optimal compression in the case where dispersion is present. If dispersion was absent, the compression factor would be $\Gamma_0 = e^{\alpha L}$. The improvement in compression due to GVD is measured by

$$D = \frac{\Gamma}{\Gamma_0} = e^{-\alpha L}\left[1 + 2\frac{\omega_0}{\omega_m}\sinh(\alpha L)\right]. \tag{74}$$

If the optimal length $L$, which is obtained as the solution of Eq. (72), is small so that $\alpha L \ll 1$, we have $\Gamma_0 = 1$ and $\Gamma = D = 1 + 2\gamma L$. If the relation $\omega_0/\omega_m \gg 1$ is well satisfied so that, despite of the condition $\alpha L \ll 1$, the parameter $\gamma L = \alpha L(\omega_0/\omega_m)$ is not small compared to unity, then the pulse compression due to GVD is substantial. The role of the modulated susceptibility in this case is limited to the frequency modulation and to the determination of the times at which compression or stretching occurs. Note that the condition $\alpha L \ll 1$ leads to $\Gamma = D \ll 1 + 2(\omega_0/\omega_m)$, that is, the compression factor $\Gamma$ is limited by the ratio $\omega_0/\omega_m$. In other words, in order to get a high compression factor, we need a large ratio $\omega_0/\omega_m$. The analysis of Ref. [13] corresponds to this situation.

If the optimal length $L$ is large so that $\alpha L \gg 1$, we find the approximation $D = \omega_0/\omega_m$. Then, in the case where $\omega_0/\omega_m \gg e^{\alpha L}$, we have $D \gg \Gamma_0$, that is, the pulse compression caused by GVD is much stronger than that caused by the time-varying susceptibility. However, because $\Gamma_0 \gg 1$ and $\Gamma = \Gamma_0 D$, both mechanisms are substantial for pulse compression in this case. In the opposite case where $\omega_0/\omega_m \ll e^{\alpha L}$, we have $\Gamma_0 \gg D$, that is, the magnitude of the pulse compression factor caused by susceptibility modulation is much stronger than the additional factor caused by GVD. This situation corresponds to the plots for the case of $z = 50$ $\mu$m in Fig. 4, where $\alpha z = 1.59$, $\Gamma_0 = 4.91$, and $D = \omega_0/\omega_m = 3.0$. For the parameters of this figure, the solution of Eq. (72) for the optimal length is $L = 52$ $\mu$m.

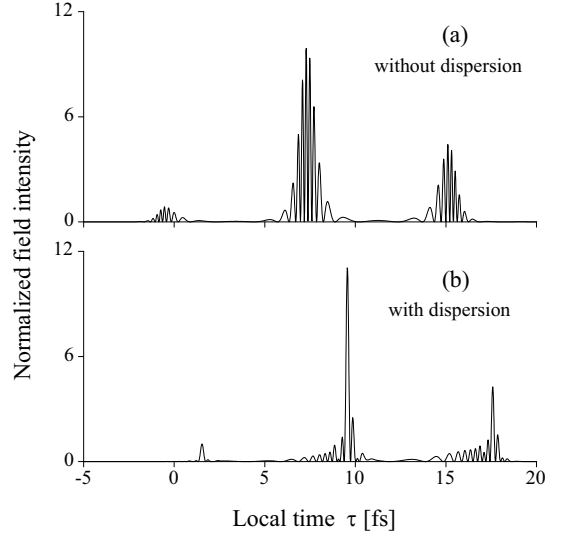

FIG. 5: Same as Fig. 4 except that for the calculations we use the input wavelength of 400 nm and the medium length of 35 $\mu$m.

The effect of GVD on pulse compression is limited by the ratio $\omega_0/\omega_m$. To illustrate a case where the ratio $\omega_0/\omega_m$ is large and, consequently, the effect of GVD on pulse compression is strong, we plot in Fig. 5 the temporal profiles of a probe pulse with a central input wavelength of 400 nm in the cases of zero (a) and nonzero (b) dispersion. We choose the medium length $z = 35$ $\mu$m ($\alpha z = 1.21$), at which the pulse compression is most profound. All the other parameters are the same as for Fig. 4. As seen, the improvement factor is on the order of $\omega_0/\omega_m = 6$, larger than the corresponding dispersionless compression factor $e^{\alpha z} = 3.34$. Note that, for the parameters of Fig. 5, the solution of Eq. (72) for the optimal length is $L = 27$ $\mu$m.

In this paper we consider the beating of a weak probe pulse with an independently prepared molecular coherence. We neglect the effect of the modulated probe pulse on the molecular states. But we can not discard the changes (produced by the molecular modulation) in the strong driving fields, which prepare the coherence. For a collinear configuration, the parameter $\alpha z$ is the same for the probe and the driving fields, and if the modulation is large enough to produce significant changes in the probe pulse, it also produces a large effect on the driving fields. Equation (38) for the phase shift per length of the prepared coherence $\rho_{ab}$ is an important assumption. It is not automatically fulfilled even when dispersion is absent. This assumption says that either the driving fields in a collinear configuration are not affected significantly ($\alpha z$ is smaller or comparable to unity) or the effect is compensated by a noncollinear beam propagation geometry.

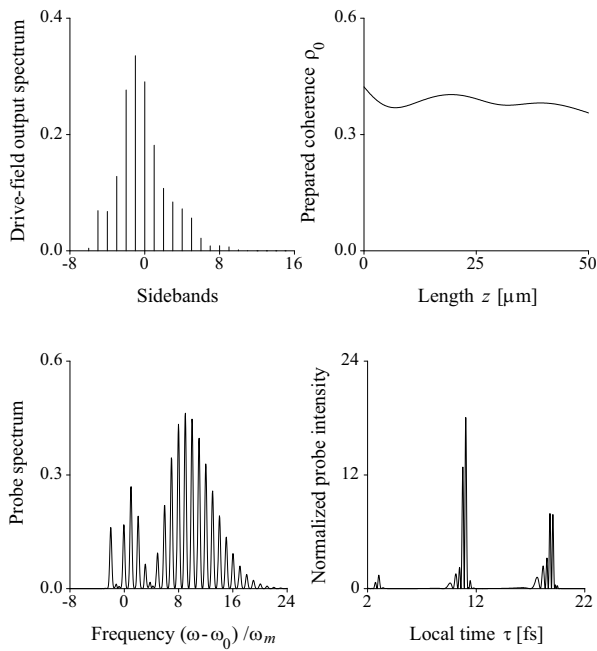

FIG. 6: Full numerical calculations for the normalized integrated spectrum of the sidebands of the driving fields, for the medium coherence prepared at the peak time $\tau_p = 0$, for the normalized probe output spectrum, and for the normalized probe output intensity. We prepare a Raman coherence in a solid hydrogen medium with a length of 50 $\mu$m by two Gaussian-shape pulse laser fields with wavelengths of 355 and 416 nm. The two driving fields have the same peak intensity of $10^9$ W/cm$^2$, the same peak time $\tau_p = 0$, and the same pulse length of 10 ns. The two-photon detuning is $\delta/2\pi = -50$ MHz. The prepared coherence modulates the frequency and compresses the length of the probe field with an input pulse length of 10 fs, an input peak time $\tau_p = 0$, and an input central wavelength of 800 nm.

There are several particular cases where self-consistent analytical solutions are possible. In the case of small dispersion and limited modulation bandwidth ($\alpha z \ll 1$), the carrier frequency is modulated but the waveform envelope is unchanged. As a result, the driving force on molecules (i.e. the two-photon Rabi frequency) remains unaffected by the sideband generation [13]. Therefore, the Bessel function solution of Ref. [13] is self-consistent. Note that in this case the number of generated sidebands ($\sim 2\gamma z$) can still be very large if $\omega_0 \gg \omega_m$. Another example of self-consistent approaches to the interaction of a light field with a molecular medium is the work on resonant non-adiabatic Raman scattering, which predicts $2\pi$ soliton formation [14, 27].

The analytical solution derived in this paper is not self-consistent, and can be used only as a tool for qualitative studies. Exact numerical simulations are required for quantitative treatments. We present in Fig. 6 the results of the full numerical calculations for the driving fields, the medium state, and the probe field. We drive a solid hydrogen medium with a length of 50 $\mu$m by two Gaussian-shape pulse laser fields at frequencies of 28,169 cm$^{-1}$ (355 nm in wavelength) and 24,019 cm$^{-1}$ (416 nm in wavelength). We choose a two-photon detuning $\delta/2\pi = -50$ MHz, at which the conditions for the adiabatic coherence preparation are satisfied and a substantial amount of molecular coherence can be produced [29]. The two driving fields have the same peak intensity of $10^9$ W/cm$^2$, the same peak time $\tau_p = 0$, and the same pulse length of 10 ns. The population and coherence decay rates of the Raman transition are $\gamma_1 = 25 \times 10^3$ s$^{-1}$ and $\gamma_2 = 10^7$ s$^{-1}$, respectively. The medium state and the sidebands of the driving fields are calculated by solving the set of the density-matrix equations (5) and the propagation equation (17). We use a probe field with an input pulse length of 10 fs, an input peak time $\tau_p = 0$, and an input central wavelength of 800 nm to beat with the molecular coherence prepared by the driving fields. The plots for the probe field in Fig. 6 show that a broad spectrum and a substantial compression of the probe pulse are produced.

## IV. CONCLUSIONS

We have studied the propagation of the field in a far-off-resonance Raman medium. We have derived the propagation equations for the field in the frequency and time domains without the use of the slowly varying envelope approximation in time. We have analyzed various aspects of the compression of a short probe pulse beating with a prepared Raman coherence. In the framework of a solvable model with zero dispersion, we have derived analytical expressions for the probe field, the oscillation frequency, and the Raman spectrum. We have shown that the modulated susceptibility determines the basic features and characteristics of the pulse compression in the Raman medium, such as the times at which compression or stretching occurs, the periodic change in the oscillation frequency, the Bessel function nature of the Raman spectrum, and the asymmetric increase of the numbers of Stokes and anti-Stokes sidebands. We have derived the conservation relations, such as the conservation of the pulse area, the conservation of the photon number, the conservation of the number of oscillations, and the conservation of the product of the pulse length and the mean frequency. We have performed numerical calculations using the parameters of solid hydrogen, and have found that the dispersion of the medium improves the pulse compression. We have demonstrated numerically the situations, where one of the two compression mechanisms, that is, the time-varying susceptibility and the GVD, is dominant compared to the other in the determination of the compression factor. The criteria for the two regimes have been derived.

In this paper we have considered excitation of a single Raman transition, such that the resultant susceptibility modulation is purely sinusoidal. Our formalism can be generalized for the case of an arbitrary number of molecular states and an arbitrarily complex modulation, with extra terms included into the propagation equations. Indeed, we can consider any complex molecular motion as a superposition of sinusoids. Because the interaction with the weak probe pulse is purely linear, contributions from different Raman transitions to the susceptibility will simply add up. For example, our formalism can be applied to describe frequency modulation and pulse compression by rotational molecular wavepackets [17, 18]. The conservation relations derived in this paper for the excitation of a single Raman transition will also be valid for more complex Raman excitations. The analysis will prove useful for generation of single subfemtosecond and sub-cycle pulses.

### Acknowledgments

The authors thank S. E. Harris, Nguyen Hong Shon, and A. K. Patnaik for helpful discussions. A. V. S. acknowledges support from Texas Advanced Research Program.

### APPENDIX A: EVOLUTION OF THE MEDIUM STATE

We call $C_j$, $C_a$, and $C_b$ the probability amplitudes of the states $j$, $a$, and $b$, respectively, in the interaction picture. We introduce the transformation

$$c_j = C_j e^{-i\omega_{ja}\tau}, \quad c_a = C_a, \quad c_b = C_b e^{-i\delta\tau}, \quad (A1)$$

where $\delta$ is the two-photon detuning and $\omega_{ik} = \omega_i - \omega_k$ the energy difference of levels $i$ and $k$. It follows from the Hamiltonian (1) and Eqs. (2) and (3) that the time evolution of the medium state is governed by the equations

$$\frac{\partial c_j}{\partial \tau} = -i\omega_{ja}c_j + \frac{i}{\hbar}E\left(\mu_{ja}c_a + \mu_{jb}c_b e^{-i\omega_m\tau}\right), \quad (A2)$$

$$\frac{\partial c_a}{\partial \tau} = \frac{i}{\hbar}\sum_j E\mu_{aj}c_j, \quad (A3)$$

$$\frac{\partial c_b}{\partial \tau} = -i\delta c_b + \frac{i}{\hbar}\sum_j E\mu_{bj}c_j e^{i\omega_m\tau}. \quad (A4)$$

Here $\omega_m = \omega_{ba} - \delta$ is the modulation frequency.

We assume that the one-photon detunings $\omega_{ja,jb} - \omega_0$ are large compared to the Rabi frequencies $\mu_{ja,jb}E_0$ as well as to the two-photon detuning $\delta$. Here, $\omega_0$ and $E_0$ are the characteristic values of the input frequency and electric field, respectively. Due to this far-off-resonance condition, the medium state changes slowly in time.

We express $c_j$ as

$$c_j = \int_{-\infty}^{\infty} d\omega\, e^{-i\omega\tau}(c_{ja\omega} + e^{-i\omega_m\tau}c_{jb\omega}), \quad (A5)$$

and assume that $c_{ja\omega}$ and $c_{jb\omega}e^{i\delta\tau}$ as well as $c_a$ and $c_b e^{i\delta\tau}$ are slowly varying functions of $\tau$. We substitute Eq. (A5) into Eq. (A2), and set the time derivatives of $c_{ja\omega}$ and $c_{jb\omega}$ equal to zero and $-i\delta c_{jb\omega}$, respectively. Then, we find

$$\begin{aligned}
c_{ja\omega} &= \frac{\mu_{ja}E_\omega c_a}{2\hbar(\omega_{ja}-\omega)}, \\
c_{jb\omega} &= \frac{\mu_{jb}E_\omega c_b}{2\hbar(\omega_{jb}-\omega)}.
\end{aligned} \quad (A6)$$

When Eqs. (A6) are introduced into Eq. (A5), we obtain

$$c_j = \frac{1}{2\hbar}\int_{-\infty}^{\infty} d\omega\, e^{-i\omega\tau}\left(\frac{\mu_{ja}c_a}{\omega_{ja}-\omega} + e^{-i\omega_m\tau}\frac{\mu_{jb}c_b}{\omega_{jb}-\omega}\right)E_\omega. \quad (A7)$$

Equation (A7) represents $c_j$ in terms of $c_a$ and $c_b$ and is justifiable insofar as the medium is far off resonance with the field.

We use Eq. (A7) to eliminate $c_j$ from Eqs. (A3) and (A4). Then, we obtain [13]

$$\frac{\partial}{\partial \tau}\begin{bmatrix} c_a \\ c_b \end{bmatrix} = i\begin{bmatrix} \Omega_{aa} & \Omega_{ab} \\ \Omega_{ba} & \Omega_{bb}-\delta \end{bmatrix}\begin{bmatrix} c_a \\ c_b \end{bmatrix}, \quad (A8)$$

where the Stark shifts $\Omega_{aa}$ and $\Omega_{bb}$ and the complex two-photon Rabi frequencies $\Omega_{ab}$ and $\Omega_{ba}$ are given by Eqs. (6).

To take into account the population decay at a rate $\gamma_1$ and the coherence decay at a rate $\gamma_2$, we use the density matrix of the medium state, which is defined by $\rho_{ik} = c_i c_k^*$. When we use Eq. (A8) and add phenomenological decay terms, we obtain the density matrix equations (5).

### APPENDIX B: PROPAGATION OF THE FIELD

The propagation of the field is governed by the wave equation

$$\left(\frac{\partial^2}{\partial z^2} - \frac{1}{c^2}\frac{\partial^2}{\partial t^2}\right)E = \mu_0\frac{\partial^2}{\partial t^2}P, \quad (B1)$$

where the polarization density $P$ is defined by

$$P = N\sum_j \langle \mu_{aj}\sigma_{aj} + \mu_{bj}\sigma_{bj} + \text{H.c.}\rangle. \quad (B2)$$

Here $N$ is the molecular density. We use the local coordinates $z$ and $\tau = t - z/c$. Equation (B1) then takes the form

$$\left(\frac{\partial^2}{\partial z^2} - \frac{2}{c}\frac{\partial^2}{\partial \tau \partial z}\right)E = \frac{1}{\epsilon_0 c^2}\frac{\partial^2}{\partial \tau^2}P. \quad (B3)$$

We make the slowly varying envelope approximation for the spatial dependence, that is, we assume that the variation of $E$ with $z$ at constant $\tau$ occurs only over distances much larger than an optical wavelength. In this case, the second-order partial derivative in $z$ can be neglected and, therefore, Eq. (B3) reduces to

$$\frac{\partial E}{\partial z} = -\frac{1}{2\epsilon_0 c}\frac{\partial P}{\partial \tau}. \qquad (B4)$$

We express the field $E$ and the polarization $P$ as Fourier integrals

$$E = \frac{1}{2}\int_{-\infty}^{\infty} e^{-i\omega\tau} E_\omega \, d\omega \qquad (B5)$$

and

$$P = \frac{1}{2}\int_{-\infty}^{\infty} e^{-i\omega\tau} P_\omega \, d\omega. \qquad (B6)$$

Then, Eq. (B4) yields

$$\frac{\partial E_\omega}{\partial z} = \frac{i\omega}{2\epsilon_0 c} P_\omega. \qquad (B7)$$

We now express the polarization density $P$, defined by Eq. (B2), in terms of the probability amplitudes $c_i$, defined by Eq. (A1). Then, we have

$$P = N\sum_j \left(\mu_{aj} c_j c_a^* + \mu_{bj} c_j c_b^* e^{i\omega_m \tau} + \text{c.c.}\right). \qquad (B8)$$

When we substitute Eq. (A7) into Eq. (B8) and use the spectral expansion (B6), we find

$$\begin{aligned}P_\omega &= 2N\hbar(a_\omega \rho_{aa} E_\omega + b_\omega \rho_{bb} E_\omega \\ &\quad + d_{\omega-\omega_m}\rho_{ba} E_{\omega-\omega_m} + d_\omega \rho_{ab} E_{\omega+\omega_m}),\end{aligned} \qquad (B9)$$

where the coefficients $a_\omega$, $b_\omega$, and $d_\omega$ are given by Eqs. (10). In deriving the above expression we have used the assumptions that the change in the medium state is slow and the two-photon detuning is small compared to the one-photon detunings. Finally, we insert Eq. (B9) into Eq. (B7). Then, we obtain the frequency-domain propagation equation (8).